# Mid-infrared pyro-resistive graphene detector on LiNbO$_3$

*Kavitha K. Gopalan[1], Davide Janner[1,2], Sebastien Nanot[1], Romain Parret[1], Mark B. Lundeberg[1], Frank H.L. Koppens[1,3]\*, Valerio Pruneri[1,3]\**

[\*]  K.K. Gopalan, Prof. D. Janner, Dr. S. Nanot, Dr. R. Parret , Dr. M. B. Lundeberg, Prof. F.H.L. Koppens, Prof. V. Pruneri
1. ICFO-Institut de Ciencies Fotoniques,The Barcelona Institute of Science and Technology, 08860 Castelldefels, Spain
E-mail: frank.koppens@icfo.eu , valerio.pruneri@icfo.eu
2. DISAT-Institute of Materials Physics and Engineering, Politecnico di Torino, IT-10129, Torino, Italy
3. ICREA-Institució Catalana de Recerca i Estudis Avançats Passeig Lluís Companys, 23,08010 Barcelona, Spain

There is great interest in developing photodetectors for infrared (IR) wavelengths, as they are essential components in many applications, e.g. vibrational spectroscopy and thermal imaging.[1] State-of-the-art IR-photodetectors (e.g. HgCdTe based) can reach high sensitivity (>10$^{10}$ cm√Hz/W) but they usually operate at very low temperatures to achieve  sufficiently low noise levels.[1] A class of IR-photodetectors that avoid cryogenic cooling is based on pyroelectricity, i.e. a property of materials that develop surface charges when subjected to temperature changes. Ferroelectric crystals, such as lithium niobate (LiNbO$_3$), lithium tantalate (LiTaO$_3$) and lead-zirconate-titanate (PZT), which possess a large spontaneous polarization, are all pyroelectric materials. When IR radiation is absorbed by the pyroelectric substrate and converted into heat, a temperature variation is produced. This in turn induces surface charges which can be measured by voltage difference across the substrate surfaces or by current through an external shunt resistor. The performance of pyroelectric based photodetectors strongly depends on the dielectric constant ($\varepsilon_r$) and loss (tan δ) of the substrate material. [2, 3] The lower $\varepsilon_r$ and tan δ, the higher the sensitivity. Unfortunately, materials with high pyroelectric coefficient typically have high $\varepsilon_r$ and/or tanδ which pose severe limitations to the photodetector

sensitivity.[2, 3]

Recently, thanks to their electrical transport and optical properties, graphene and other 2-D materials have been investigated as promising candidates for IR photodetection.[4-8] More specifically, graphene can be combined with ferroelectric crystal substrates, where changes in spontaneous polarization can produce strong doping in graphene.[9-12] In one of the first attempts, Hsieh et al. reported an opto-thermal field effect transistor (FET) using graphene on PZT substrate, where the drain current was modulated by a near IR laser (1064 nm), with an amplitude of $3.6 \times 10^{-4}$ A/W.[13] The achieved quasi-DC modulation of the current through light absorption was measured with mm-sized graphene devices and was related to changes in graphene resistance through pyroelectric charges on PZT surface. Although revealing the opto-thermal effect, the frequency response and a full description of the mechanism to be used in photodetectors were not addressed.

More recently, an IR-photodetector based on graphene on polyvinylidene fluoride (PVDF) substrate was reported by Kulkarni et al. [14] In this case, graphene was used as transparent electrode in a classical pyroelectric detector configuration, exploiting its transparency in the IR region. Baeumer et al. instead reported spatial carrier density modulation in graphene on periodically poled $LiNbO_3$.[15] Taking advantage of the fact that charges of different signs are induced in oppositely oriented domains, they were able to demonstrate a p-n junction photodetector by applying a gate voltage across the domain inverted structure, with a responsivity of $2 \times 10^{-5}$ A/W in the visible range.

Being chemically stable and having a high Curie temperature (1415 K), $LiNbO_3$ has been widely used to make pyroelectric detectors and is compatible with the processing steps required for graphene device fabrication. Here we propose and investigate a photodetector in the mid-IR (6-10µm wavelength region) using graphene on $LiNbO_3$, which leverages the high absorption of $LiNbO_3$ and doping sensitivity of graphene. [16-18]

The basic principle of operation is shown in **Fig. 1a**. The impinging light is absorbed by the substrate and subsequently converted into heat, resulting in a local temperature increase. This produces a variation of the LiNbO$_3$ spontaneous polarization due to the pyroelectric effect. The LiNbO$_3$ bound (polarization) surface charges induce (capacitively) free charges into graphene. Through this process, the light induces a change of the electrical resistance (conductance) of the graphene channel. This phenomenon can be called pyro-resistive effect and can be exploited to obtain a new generation of photo-detectors. Indeed, the characteristics of such detectors lay in between pyroelectric ones and bolometers as they use pyroelectricity of the substrate to induce doping (as in pyro-detectors) and the read-out is based on resistance change in graphene (as in bolometers). The proposed approach can potentially lead to an optimum trade-off of both types of detectors, overcoming their intrinsic limitations when taken singularly.

Indeed, we will show that, through a proper optimization, the performance of the pyro-resistive devices could reach those of aforementioned state-of-the-art IR photodetectors, paving the way to low cost, low power consumption, graphene based photodetectors with high detectivity over a broad mid-IR range.

*Pyroelectrically induced carrier modulation at the graphene/LiNbO$_3$ interface*

In order to study the pyroelectrically induced doping, the sheet resistance of graphene deposited on LiNbO$_3$ was measured upon temperature variation.[13] Measurements were taken using 4-points configuration in a Hall bar geometry (see Methods) while sweeping the temperature of the chip across few tens of °C in vacuum (P≤10$^{-5}$ mbar) in a closed cycle helium cryostat. The resistance of the devices was monitored by using a lock-in amplifier and measuring the voltage drop across the graphene Hall bar, while injecting 1 µA AC current at 503 Hz. As shown in **Fig. 1c**, temperature variations induced a change in resistance moving across the charge neutrality point (CNP) and the typical Lorentzian shape associated with field effect was observed. A more detailed analysis (see SI.1) quantitatively demonstrates the effect of doping obtained by pyroelectric charges and shows very good agreement with the experimental data.

To study the conversion of optical signals into a variation of the graphene channel resistance, i.e. a pyro-resistive photodetector, we tested the photoresponse of several devices, which had a two-point probe geometry (**Fig. 1b** and Methods). The photoresponse measurements, which included spatial mapping, were taken by illuminating the device with a tightly focused light beam from a quantum cascade laser (QCL) operating in the 1000-1600 cm$^{-1}$ range. A mechanical chopper allowed lock-in detection while a source-drain DC voltage of 0.1V was applied to the graphene devices across the metallic side contacts. This configuration permitted to resolve the photoresponsivity not only spatially, but also with respect to the excitation frequency. The photoresponsivity maps (**Fig. 1d**) clearly show response from the regions covered with graphene both in direct-current (DC) and alternate-current (AC) at 77Hz. The presence of photoresponse signal from regions outside the graphene area is due to lateral heat propagation in the substrate. Little or no signal is detected at the contacts as the gold reflects most of the radiation and thus no significant heating of the substrate occurs.

*Model of pyro-resistive photodetection*

The pyro-resistive photodetector can be described with a simple model that helps to get a physical insight and can be eventually used to optimize the design for best performance. Indeed, for a layer of graphene deposited on top of a pyroelectric (ferroelectric) substrate (e.g. z-cut LiNbO$_3$), the electrical resistance (*R*) change against optical power (*P*) can be expressed as (see SI for details):

$$\frac{dR}{dP} = N_{EQ} \frac{dR_S}{dn} \left[\frac{dq_{IND}}{dn}\right]^{-1} \frac{dq_{IND}}{dT} \frac{dT}{dP} \qquad (1)$$

Here, N$_{EQ}$ is the number of equivalents of sheet resistance (*R$_s$*) of the device and depends on the geometry and patterning of graphene, *dR$_s$/dn* is the variation of sheet resistance with respect to carrier density and depends on graphene properties such as mobility, intrinsic doping (*n*) and Fermi energy, $E_f = hv_f\sqrt{n\pi}$. The surface pyroelectric charge is represented by $q_{IND}$ and the term $(dq_{IND}/dn)^{-1}$ accounts for the number of free carriers in graphene produced by each pyroelectric induced charge in the substrate and depends on the density of states in graphene (typically is set to *1/e*). The factor $dq_{IND}/dT$ is proportional to the pyroelectric coefficient of the substrate and *dT/dP* is the change in temperature induced by the incident optical power, which depends on optical absorption and thermal conductivity of the substrate.

The different terms in Eq. (1) point out clear design guidelines. Indeed, three main contributions can be identified: device geometry ($N_{EQ}$), graphene quality and properties ($dR_s/dn$ $(dq_{IND}/dn)^{-1}$) and substrate material ($dq_{IND}/dT \cdot dT/dP$). All these contributions play a crucial role in the optimization of performance. For example, a straightforward approach in order to increase *dR/dP* could be to leverage the geometry, e.g. use of a serpentine, to increase $N_{EQ}$. However, to better illustrate the role of each factor contributing to the pyro-resistive photodetection, in the following, we will focus on graphene's quality, surface properties and substrate geometry.

*Photoresponse of pyro-resistive graphene devices*

Since, according to model in Eq. (1), the photoresponse intensity is expected to be dependent on the Fermi level of graphene, measurements were taken at different graphene doping levels. The change of doping is like a bias to tune the working point of the pyro-resistive detector and was achieved by top-gating the devices using a polymer ion gel (LiClO$_4$: Poly ethylene oxide (PEO)). Corresponding results are shown in **Fig. 2**. Before testing the photoresponse of the graphene device, the sheet resistance dependence on top-gating voltage was measured in order to extract the physical parameters of graphene. As shown in **Fig. 2a,** a Lorentzian model for graphene conductivity (σ) against the top-gating voltage closely fits the experimental data with $n_0$=7.3 x 10$^{12}$. After electrical characterization, the photoresponsivity curve defined as ΔI/I was measured versus the top-gating voltage (**Fig. 2b**). The curve follows the first derivative of log (σ) (see SI 4.2) and the fitting is in very good agreement with the experimental results for the n-doped region. In the p-doped region the deviation from the Lorentzian shape (Fig. 2a and 2b) can be attributed to the ion-gel, which is known to affect graphene mobility.[19] Note that the deviation from the ideal behavior occurs both for resistance and photoresponse. The point of maximum photoresponsivity in Fig. 2b, which is of interest for photodetection applications, corresponds to the flex in the Lorentzian curve of Fig. 2a as expected from the analysis in S.4. The dependence of the photoresponsivity in the proposed photodetector can be used not only to achieve the highest response by tuning the bias but also to reduce the sensitivity of the device when measuring intense light. The possibility to tune the sensitivity increases its dynamic range of several orders of magnitude, this being an essential feature in many applications.

Another fundamental characteristic of photodetectors is their temporal response which is directly related to their bandwidth and give constraints on their potential applications. The response of a mid-IR photodetector is both related to its electronic response and the heat propagation in the device[1]. To characterize the temporal response and the heat conduction of the pyro-resistive photodetector, frequency response measurements (**Fig. 3a**) were carried out by varying the laser chopper frequency at the point of maximum responsivity determined by the maps (Fig.1d and SI.5). In comparison with conventional pyroelectric detectors that are not responsive at low frequencies, the pyro-resistive device shows photoresponse even in DC. The photoresponse normalized to the DC value shows a decay with frequency that can be accounted for by a phenomenological model. The photodetector system response $H(\omega)$ can be approximated as (see SI for details):

$$|H(\boldsymbol{\omega})| = \frac{1}{\sqrt{1+\tau^2\omega^2}}|\hat{F}(\omega)| \qquad (2)$$

where the first term $1/\sqrt{(1+\tau^2\omega^2)}$ in Eq. 2 is the classical behavior of a thermal detector[1] with time constant (τ) and $\hat{F}(\omega)$ is a rational polynomial term that comes mainly from heat propagation characteristics in the bulk LiNbO$_3$ around the device and the dissipation of heat through air (see SI). As shown in **Fig. 3b**, the fitting according to Eq. 2 is in good agreement with the experimental data and gives $\tau$ =1.3 s (see S3 for more details). τ is mainly related to the thermal behavior of the thick LiNbO$_3$ substrate and could be reduced strongly by thinning the substrate, i.e. reducing its heat capacitance.

From the measurements performed on the device, we could also extract the thermal (heat propagation) length in the graphene/LiNbO$_3$ device by the photoresponse maps at different frequencies (**Fig. S3**). Taking cross sections of the maps along the y axis at the maximum values of the photoresponse we obtained the behavior of thermal propagation with distance (see SI 8 and Fig. S8.1). The thermal length is identified by the half-width-at-half maximum of the gaussian photoresponse distribution that follows (in the short range) the temperature profile and is proportional to $f^{-1/4}$ as shown in Fig. 3b. Indeed, assuming the temperature profile to have a gaussian shape of the form $\exp[-(r/w)^2]$, $w$ depends on the chopping frequency as $w(f)=\beta f^{-1/4}$ with β=146 μm/Hz$^{1/4}$. This means that the higher the frequency the shorter the time that the heat generated by the laser has to dissipate radially and along the thickness of the substrate. Such behavior is the result of heat dissipation through conduction in the LiNbO$_3$ substrate and air convection at the surface, similar to Logan *et al*.[20]

The maximum detectivity (D$^*$) obtained for the tested pyro-resistive photodetector was 1.14 ×10$^5$ cm√Hz/W (for details on this estimate one can refer to SI.3). This value of detectivity can be strongly increased by thinning the LiNbO$_3$, i.e. reducing heat conduction through the substrate, and using an environment less conductive than air.[21] For example, if a thin film of a mid-IR absorber is deposited on LiNbO$_3$ with a smaller thickness (e.g. 50 μm) and the device is operated in vacuum – as is the case of common pyroelectric detectors - we can expect a reduction of several orders of magnitude for the time constant, to ms levels and an increase of the local temperature raise by laser irradiation with a proportional increase in the detectivity of the pyro-resistive photodetector (see **Fig. S6** and analysis in the SI). Such improvement combined with the use of a clean graphene with $n_0 \sim 10^{11}$ opens up the possibility of reaching detectivity levels of the order of $\sim 10^8$-$10^{10}$ cm√Hz/W, which are very high for uncooled detectors operating in the mid-IR.

*Conclusions*

In conclusion, we have demonstrated that graphene in combination with $LiNbO_3$ crystal substrates can be used to make efficient photodetectors in the mid-IR based on pyro-resistive effect, which do not require cryogenic cooling. The demonstrated detectivity compares well with previous graphene based devices and can be increased by an order of magnitude by reducing the initial doping and more than two orders of magnitude by employing a thin sheet of $LiNbO_3$ with an absorber on top (as detailed in the SI). Operating the device in vacuum would reduce the heat dissipation, eventually limited to radiation effects, and thus increase the photoresponse even further.

*Methods:*

The tested devices consisted of hall bar and two probe geometries fabricated on Z-cut $LiNbO_3$ with a gold back contact. All the graphene devices were fabricated on z-cut $LiNbO_3$ using UV lithography resist patterning, subsequent thermal evaporation of Cr/Au contacts and graphene sheets grown via chemical vapor deposition (CVD) were transferred using a wet process (see SI for more details on fabrication).[22] Etching and patterning of graphene was carried out with $O_2$/Ar plasma at a power of 10 W for about 1 minute.

*Hall bars* for studying the pyroelectrically induced doping had a length of 50 µm and a width of 40 µm. The as-deposited sheet resistance was 4130 Ω, and the mobility was about 4600 $cm^2$/Vs for the presented device.

*Two-point probe photodetector* devices had different sizes of few tens of microns in a rectangular shape. The graphene mobility of the fabricated devices typically ranged from 500 to 1500 cm$^2$/Vs and the residual initial doping corresponded to a Fermi level of about -0.2eV. These values were obtained by Raman spectroscopy and Hall measurements characterization (see SI). [23, 24]

*Top gating* of the photodetector devices was done by drop-cast ion gel polymer (LiClO$_4$: Poly ethylene oxide (PEO)). [25,26] From top gating measurements we could estimate a contact resistance of 5kΩ. The same ion-gel doping was used as a way to bias the photodetector and test the photoresponse for different initial doping levels.

*Acknowledgement*

We acknowledge financial support from AGAUR (2014 SGR 1623) and the Spanish Ministry of Economy and Competitiveness (MINECO), through the "Severo Ochoa" Programme for Centres of Excellence in R&D (SEV-2015-0522), the "Fondo Europeo de Desarrollo Regional" (FEDER) through grant TEC2013-46168-R and from the European Union H2020 Programme under grant agreement (no. 696656) "Graphene Flagship". K. K. G. acknowledges the International PhD fellowship "la Caixa" - Severo Ochoa @ ICFO.

**References**

[1]　A. Rogalski, *Infrared detectors*, CRC press, 2010.
[2]　M. Paul, Reports on Progress in Physics 2001, 64, 1339.
[3]　R. W. Whatmore, Reports on Progress in Physics 1986, 49, 1335.
[4]　T. Mueller, F. Xia, P. Avouris, Nat Photon 2010, 4, 297.
[5]　G. Konstantatos, M. Badioli, L. Gaudreau, J. Osmond, M. Bernechea, F. P. G. de Arquer, F. Gatti, F. H. L. Koppens, Nat Nano 2012, 7, 363.
[6]　F. Bonaccorso, Z. Sun, T. Hasan, A. C. Ferrari, Nat Photon 2010, 4, 611.
[7]　M. Badioli, A. Woessner, K. J. Tielrooij, S. Nanot, G. Navickaite, T. Stauber, F. J. García de Abajo, F. H. L. Koppens, Nano Letters 2014, 14, 6374.
[8]　A. L. Hsu, P. K. Herring, N. M. Gabor, S. Ha, Y. C. Shin, Y. Song, M. Chin, M. Dubey, A. P. Chandrakasan, J. Kong, P. Jarillo-Herrero, T. Palacios, Nano Letters 2015, 15, 7211.
[9]　E. B. Song, B. Lian, S. Min Kim, S. Lee, T.-K. Chung, M. Wang, C. Zeng, G. Xu, K. Wong, Y. Zhou, H. I. Rasool, D. H. Seo, H.-J. Chung, J. Heo, S. Seo, K. L. Wang, Applied Physics Letters 2011, 99, 042109.
[10]　G.-X. Ni, Y. Zheng, S. Bae, C. Y. Tan, O. Kahya, J. Wu, B. H. Hong, K. Yao, B. Özyilmaz, ACS Nano 2012, 6, 3935.
[11]　D. Jin, A. Kumar, K. Hung Fung, J. Xu, N. X. Fang, Applied Physics Letters 2013, 102, 201118.
[12]　Y. Zheng, G.-X. Ni, C.-T. Toh, M.-G. Zeng, S.-T. Chen, K. Yao, B. Özyilmaz, Applied Physics Letters 2009, 94, 163505.
[13]　C.-Y. Hsieh, Y.-T. Chen, W.-J. Tan, Y.-F. Chen, W. Y. Shih, W.-H. Shih, Applied Physics Letters 2012, 100, 113507.
[14]　E. S. Kulkarni, S. P. Heussler, A. V. Stier, I. Martin-Fernandez, H. Andersen, C.-T. Toh, B. Özyilmaz, Advanced Optical Materials 2015, 3, 34.
[15]　C. Baeumer, D. Saldana-Greco, J. M. P. Martirez, A. M. Rappe, M. Shim, L. W. Martin, Nat Commun 2015, 6.
[16]　Z. Q. Li, E. A. Henriksen, Z. Jiang, Z. Hao, M. C. Martin, P. Kim, H. L. Stormer, D. N. Basov, Nat Phys 2008, 4, 532.
[17]　C. Marín, A. G. Ostrogorsky, G. Foulon, D. Jundt, S. Motakef, Applied Physics Letters 2001, 78, 1379.
[18]　Y. V. Shaldin, V. Gabriélyan, S. Matyjasik, Crystallography Reports 2008, 53, 847.
[19]　B. J. Kim, H. Jang, S.-K. Lee, B. H. Hong, J.-H. Ahn, J. H. Cho, Nano Letters 2010, 10, 3464.
[20]　R. M. Logan, K. Moore, Infrared Physics 1973, 13, 37.
[21]　P. Muralt, Reports on Progress in Physics 2001, 64, 1339.
[22]　A. Reina, H. Son, L. Jiao, B. Fan, M. S. Dresselhaus, Z. Liu, J. Kong, The Journal of Physical Chemistry C 2008, 112, 17741.
[23]　A. C. Ferrari, D. M. Basko, Nat Nano 2013, 8, 235.
[24]　K. S. Novoselov, A. K. Geim, S. V. Morozov, D. Jiang, M. I. Katsnelson, I. V. Grigorieva, S. V. Dubonos, A. A. Firsov, Nature 2005, 438, 197.
[25]　G. P. Siddons, D. Merchin, J. H. Back, J. K. Jeong, M. Shim, Nano Letters 2004, 4, 927.
[26]　C.-F. Chen, C.-H. Park, B. W. Boudouris, J. Horng, B. Geng, C. Girit, A. Zettl, M. F. Crommie, R. A. Segalman, S. G. Louie, F. Wang, Nature 2011, 471, 617.

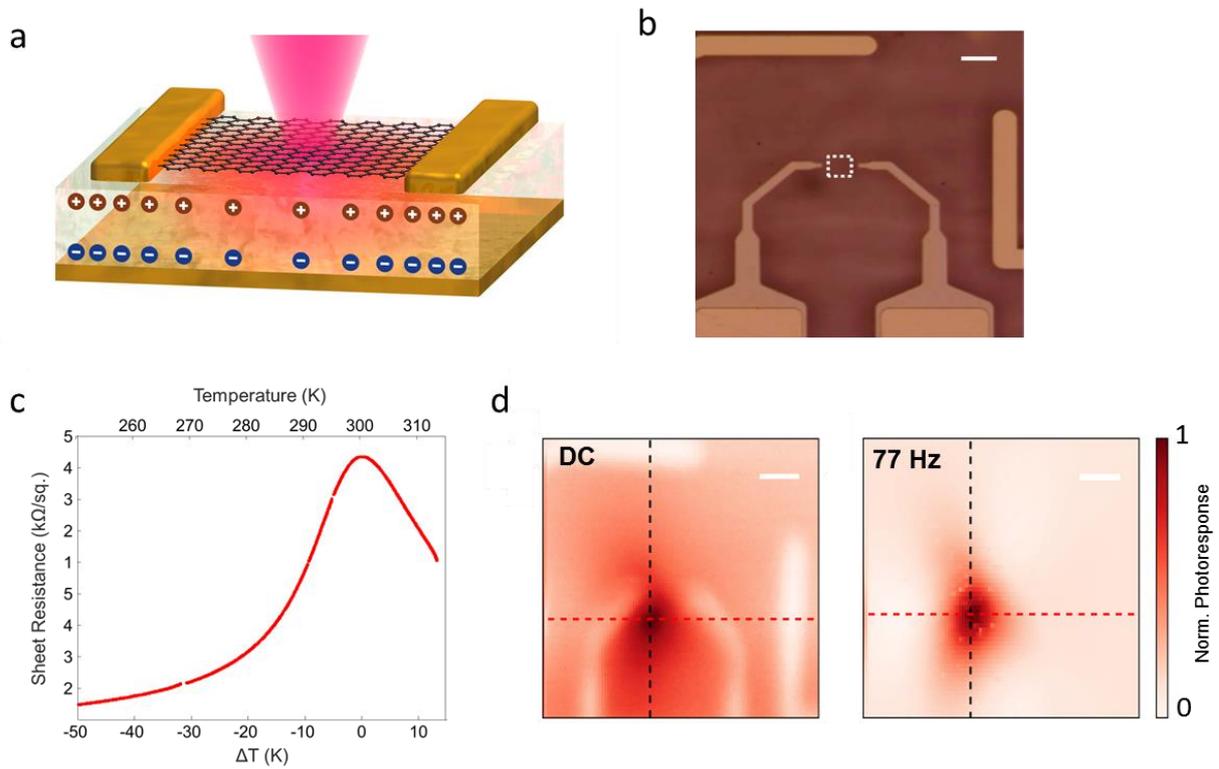

**Figure 1.** a) Schematic representation of the Graphene/LiNbO$_3$ photodetection device. Mid-IR radiation (1000-1600 cm-1) is shown in pink, the positive and negative surface charges in brown and blue, respectively. Under mid-IR radiation absorption, an increase in temperature occurs leading to a decrease in surface charge density in LiNbO$_3$. This in turn induces doping into the graphene layer and change of its resistivity. b) Optical microscope image of the graphene/ LiNbO$_3$ device. Graphene region is marked by the square. Scale bar is 100 µm. c) Pyroelectrically induced doping and resistance change in graphene on LiNbO$_3$ during cooling (heating) in a closed cycle helium cryostat. The curve sweeps across the charge neutrality point (CNP) without the need of an external gate voltage. d) Normalized photoresponse maps measured in DC and at 77 Hz . The maps were obtained by scanning on the device a focussed laser beam of wavelength λ =9.26 µm and power P = 1.8 mW. Scale bar is 100 µm.

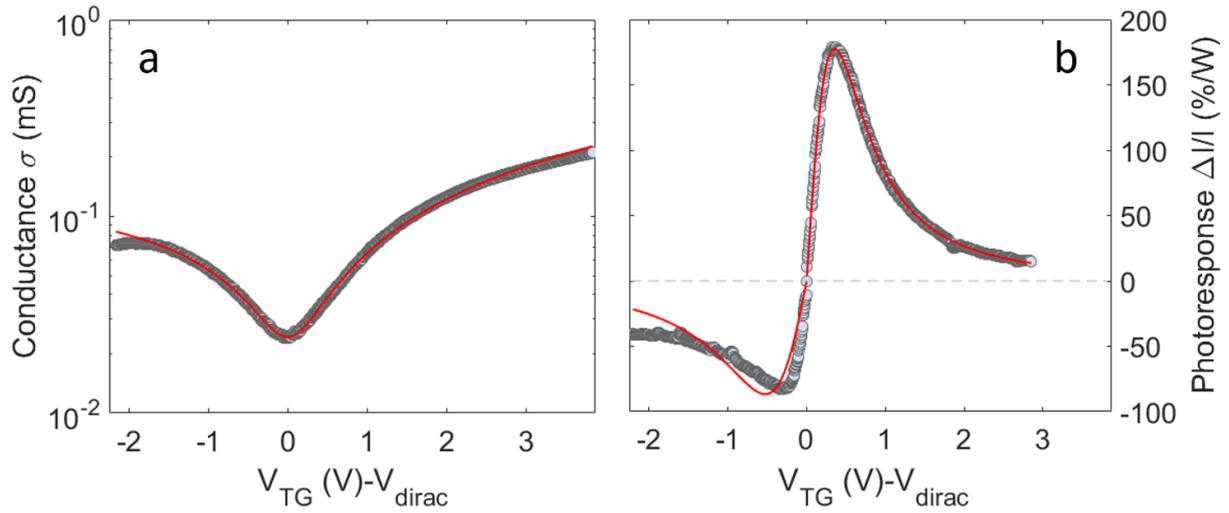

**Figure 2.** a) Conductance of graphene on LiNbO$_3$ obtained by top gating with ion gel. b) Photoresponse dependence with top gating voltage. The photoresponse defined as ΔI/I per Watt follows the derivative of log (σ) and can be fitted as described in the text. The best fit with such function on the negative branch is not accurate, since the same Dirac curve deviates significantly from the lorentzian behavior.

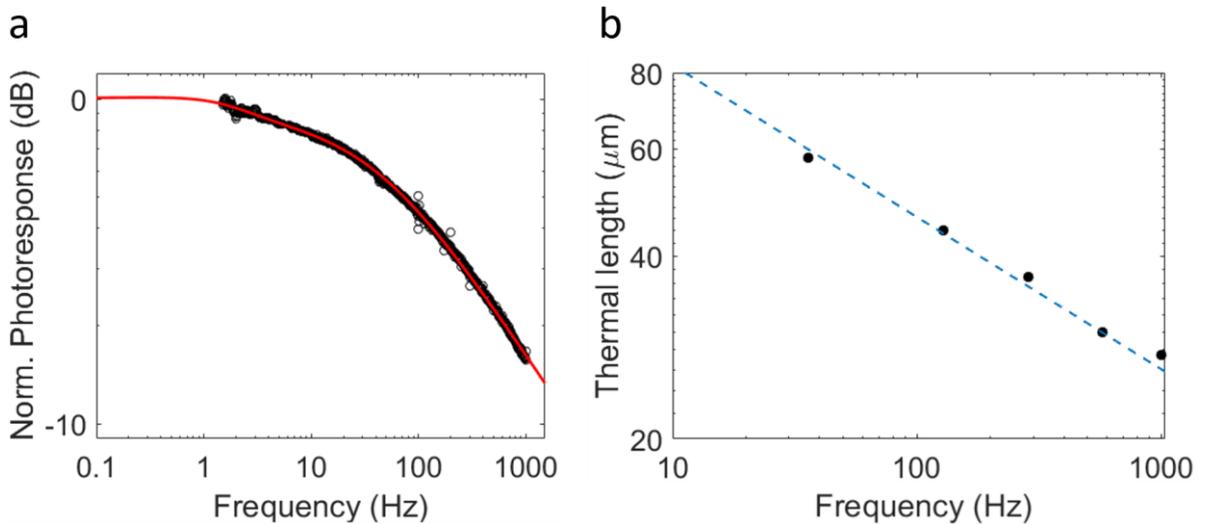

**Figure 3.** a) Dependence of the normalized responsivity with respect to frequency (black dots). The fitting (red solid line) is obtained with Eq.3. b) Dependence of the thermal length with frequency.

**Supplementary Information**

**Mid-infrared pyro-resistive graphene detector on LiNbO$_3$**

*Kavitha K. Gopalan[1], Davide Janner[1,2], Sebastien Nanot[1], Romain Parret[1], Mark Lundeberg[1], Frank H.L. Koppens[1], Valerio Pruneri[1,3]\**

*S.1 Characterization of pyro-resistive effect*

The pyro-resistive behavior of graphene on LiNbO$_3$ was characterized by measuring devices with Hall bar geometry, with a length of 50 µm and a width of 40 µm. Measurements were taken while sweeping the temperature of the chip across few tens of ºC in vacuum (P≤10$^{-5}$ mbar) in a closed cycle helium cryostat. The measurements are shown in figure 1c in the main text. Below in figure S1 we show the same figure where we have converted the variation of temperature with respect to the charge neutrality point ($\Delta T$) into the carrier density induced in graphene ($n$), in contact with LiNbO$_3$. The conversion is readily obtained via the relation:

$$n = \frac{\gamma(T)}{e} \Delta T \qquad \text{SI.1}$$

Where *e* is the electron charge and $\square(T)$ is the pyroelectric coefficient at temperature $T$ [1].

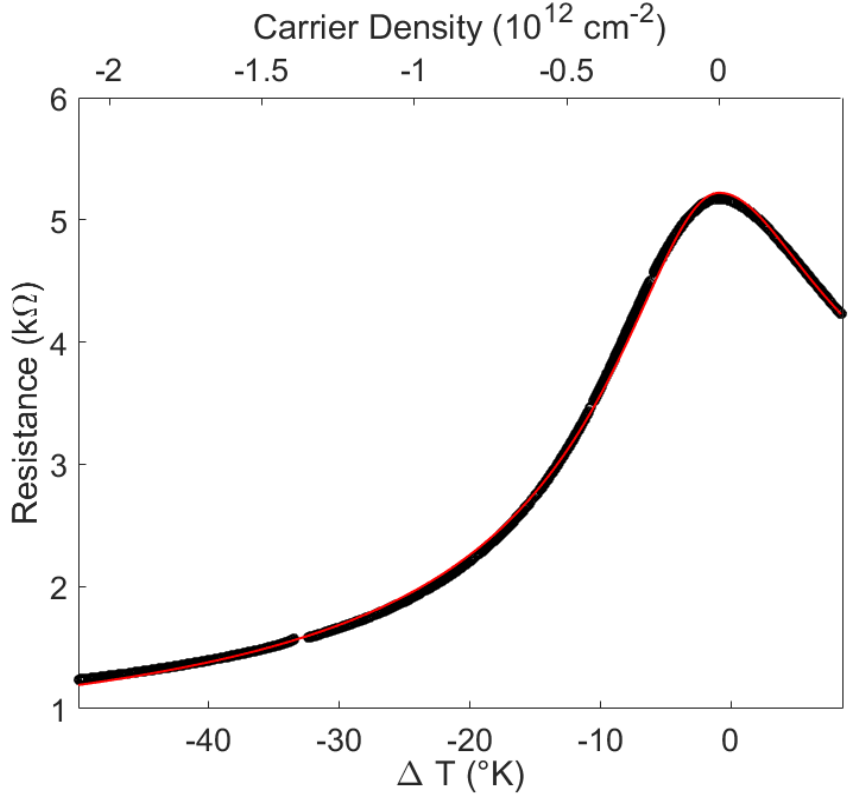

**Figure S1** Temperature dependence of the resistance of the Hall bar graphene device. Black dots indicate the measured values and the continuous red line is the Lorentzian fitting following the Dirac curve model of a field-effect graphene device.

The fitting of the curve was performed with the Lorentzian model for the resistivity of graphene as a function of the carrier density:

$$R(n) = R_c + N_{EQ} \frac{1}{e\mu\sqrt{n_0^2+n^2}} \qquad \text{SI.2}$$

where $R_c$ is the contact resistance of the device, $N_{EQ}$ the number of equivalents (given the shape $N_{EQ}$=1.25), m the mobility and $n_0$ is related to the intrinsic impurities in graphene. The parameters for the curve fitted in figure S1 are: $R_c$=440Ω, μ=4639 cm²/Vs, $n_0$=3.5 $10^{11}$cm⁻².

### S.2 Raman characterization

Raman measurements were taken with a 532 nm laser focused to a spot diameter of 1 μm (FWHM) onto the graphene layer using a 50x objective. The signal was accumulated for 10 seconds. Figure S2 shows the characteristic graphene-Raman peaks G and 2D at 1587 cm$^{-1}$ and 2700 cm$^{-1}$, respectively.

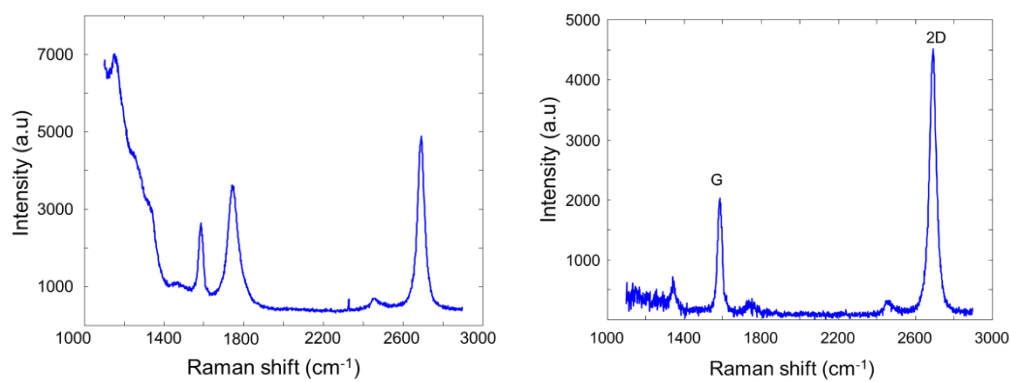

**Figure S2:** a) Raman measurements on the graphene on LiNbO$_3$ device with gold back contact b) Raman measurements after subtraction of the LiNbO$_3$ /gold background

*S.3 Detectivity and Wavelength dependence of the responsivity*

The responsivity of the graphene/LiNbO$_3$ photodetector is dependent on the absorption of the impinging laser radiation in the substrate. To verify such behavior we measured the photoresponse of the device in the point of maximum responsivity both in DC and at 77Hz and compared it with the absorption. In figure S1 we show such comparison. The DC measurements present higher fluctuations, mainly related to the fact that their detection is more complexity while those at 77Hz follow the wavelength absorption dependence of LiNbO$_3$.

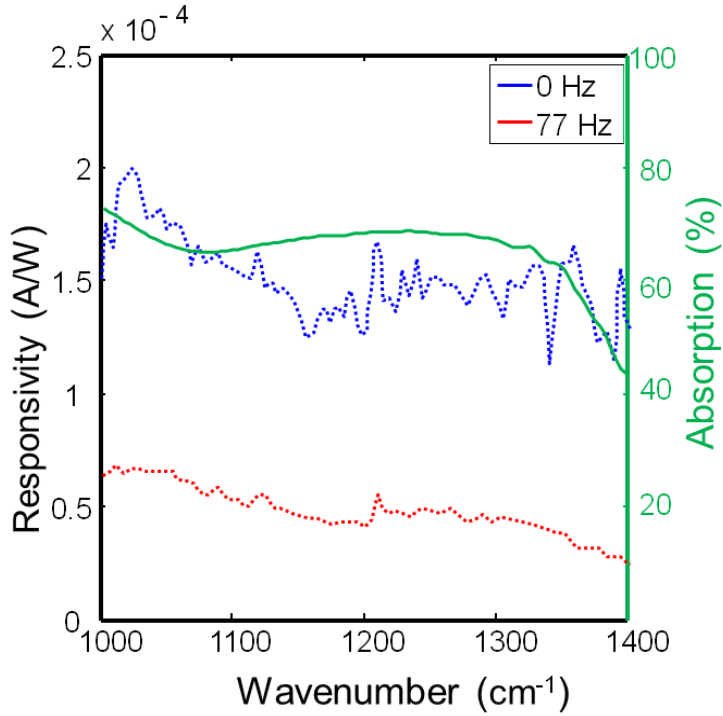

**Figure S3.1** Wavenumber dependence of the absorption in LiNbO$_3$ for a 500um thick substrate (green) and the responsivity of the photodetector measured in DC (blue dotted) and AC at 77 Hz (red dotted)

As shown in figure S3.2, noise spectral density measurements were also done on the devices and the detectivity was calculated as:

$$D^* = \frac{Responsivity \times \sqrt{Area}}{S_n}$$

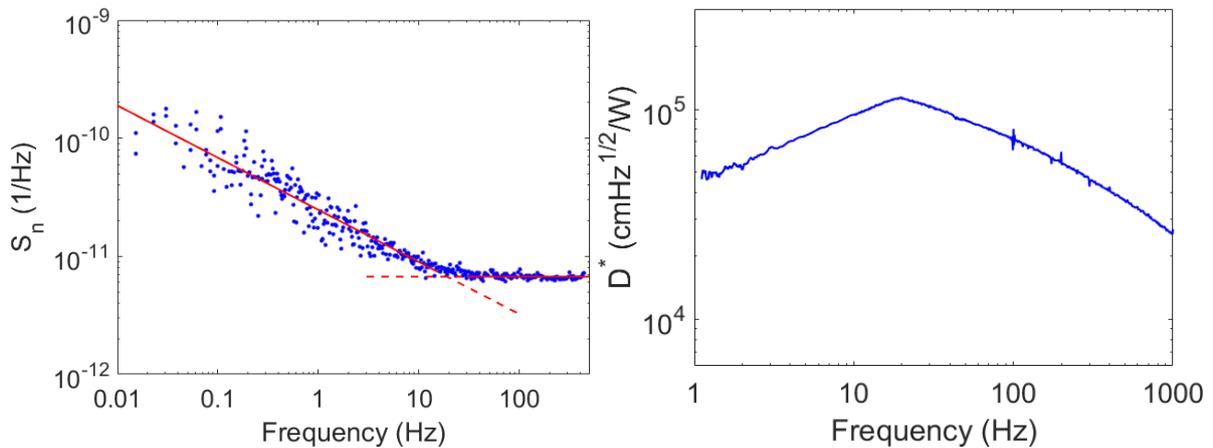

**Figure S3.2** Noise spectral density and detectivity D$^*$ of the device in S3.1

In order to calculate the detectivity $D^*$ we used the fitted values for the noise $S_n$. In figure S3.2 we plot the detectivity for frequencies up to 1000Hz from the measured photoresponse. At 20 Hz, for the tested device with an area of 75µm×75µm we reached the maximum detectivity $D^* = 1.14 \times 10^5 \mathrm{cm}\sqrt{Hz}/W$.

*S.4 Pyro-resistive graphene photodetectors response analysis*

The pyroelectric doping effect can be described with a simple model that can also be used to optimize pyro-resistive devices. For a layer of graphene deposited on top of a z-cut LiNbO$_3$ substrate, the electrical resistance (R) change can be expressed as:

$$\frac{dR}{dP} = \frac{dR}{dq_{IND}} \frac{dq_{IND}}{dT} \frac{dT}{dP} \qquad \text{SI.3}$$

Where $q_{IND}$ is the surface pyroelectric charge and dT/dP the change in temperature induced by the incident optical power P, which depends on optical absorption and thermal conductivity of the LiNbO$_3$.

The previous expression can be expanded as:

$$\frac{dR}{dP} = N_{EQ} \frac{dR_s}{dn} \left[\frac{dq_{IND}}{dn}\right]^{-1} \frac{dq_{IND}}{dT} \frac{dT}{dP} \qquad \text{SI.4}$$

Where $N_{EQ}$ is the number of equivalents of sheet resistance ($R_s$) of the device and depends on the geometry and patterning of graphene, $dR_s/dn$ is the variation of sheet resistance with respect to carrier density and depends on graphene properties such as mobility, intrinsic doping (n) and Fermi energy, $E_f = hv_f\sqrt{n\pi}$. The term $(dq_{IND}/dn)^{-1}$ accounts for the number of carriers produced by each pyroelectric induced charge and depends on the density of states in graphene (typically is set to 1/e). The factor $dq_{IND}/dT$ is proportional to the pyroelectric coefficient of the substrate. In the simplest case, we can rewrite:

$$\frac{dR}{dP} = \frac{N_{eq}\gamma_{pyro}\eta_{Opt-Th}}{e} \frac{dR_s}{dn} \qquad \text{SI.5}$$

where $\gamma_{pyro}$ is the pyroelectric coefficient of the substrate and $\eta_{Opt-Th}$ accounts for the conversion efficiency from optical power to thermal power in the substrate.

*S.4.1 Parameters of interest in performance of pyro-resistive detectors*

In eq. SI5, the term mostly related to graphene porperties is $dR_s/dn$. If, for the sake of simplicity, we assume for the sheet resistance $R_s$ a dependence with respect to n as:

$$R_S = \frac{1}{e\mu n_0} \frac{1}{\sqrt{1+\frac{n^2}{n_0^2}}} \qquad \text{SI.6}$$

the derivative of $R_s$ in eq. SI.5 is:

$$\frac{dR_s}{dn} = -\frac{n}{e n_0^3 \mu \left(1+\frac{n^2}{n_0^2}\right)^{3/2}} \qquad \text{SI.7}$$

which has a maximum for $|n| = n_0/\sqrt{2}$. At its maximum we have

$$\left.\frac{dR_s}{dn}\right|_{n=n_0/\sqrt{2}} = -\frac{2}{3\sqrt{3} e n_0^2 \mu} \qquad \text{SI.8}$$

Considering eq. SI.6 and the fact that $max(R_s) = 1/e\mu n_0$, we can thus assume that the maximum value of $dR_s/dn$ is limited by the quantum of conductance such that $max(en_0\mu) = \frac{4e^2}{h}$ which substituted in the previous equation reads:

$$\left.\frac{dR_s}{dn}\right|_{n=n_0/\sqrt{2}} \leq -\frac{h}{2e^2 n_0} \qquad \text{SI.9}$$

Within this approximation, for the best graphene samples the intrinsic impurities density $n_0$ limits the responsivity.

Although it is meant mainly for materials that have an exponential dependence of the resistance change with respect to temperature, a similar conclusion can be drawn if instead of the full expansion of SI.5 we consider the widespread parameter of the Temperature Coefficient of Resistance (TCR). The TCR defined as α =1/R (dR/dT) for pyro-resistive graphene devices is:

$$\alpha_{TCR} = \frac{1}{R_s}\frac{dR}{dn} \qquad \text{SI.10}$$

The maximum of the TCR occurs for $n=n_0$:

$$max(\alpha_{TCR}) = \frac{1}{2n_0} \qquad \text{SI.11}$$

thus further confirming the dependence of the responsivity with respect to the intrinsic impurity density of graphene.

*S.4.2 Characterization of photoresponse at different graphene doping*

The photoresponse measurements performed on the devices were realized by using a coherent detection technique by chopping the laser beam at a frequency of 77 Hz and detecting the current generated by the variation of the resistance in graphene at that frequency while keeping the source to drain voltage across it constant ($V_{bias}$). In this case, the light impinging on the LiNbO$_3$ substrate generate an increase in temperature that is proportional to the optical power as $\Delta T = \eta_{Opt-Th} P$. The modulation of graphene carrier density produced by the pyro-electric effect can be written as $\Delta n = \frac{\gamma_{pyro}}{e}\Delta T$, and the corresponding change in resistance can be written as $\Delta R = \frac{dR}{dn}\Delta n$. Combining these expressions and after some calculation we can express the modulated current produced by the light as:

$$\Delta I = -\frac{V_{bias}}{R^2}\frac{dR}{dn}\frac{\gamma_{pyro}}{e}\eta_{Opt-Th}P \qquad \text{SI.12}$$

From this equation directly follows that □I depends on d□/dn and that □I/I depends on dlog(□)/dn where □=1/R is the graphene conductivity.

*S.5 Responsivity maps at different frequencies*

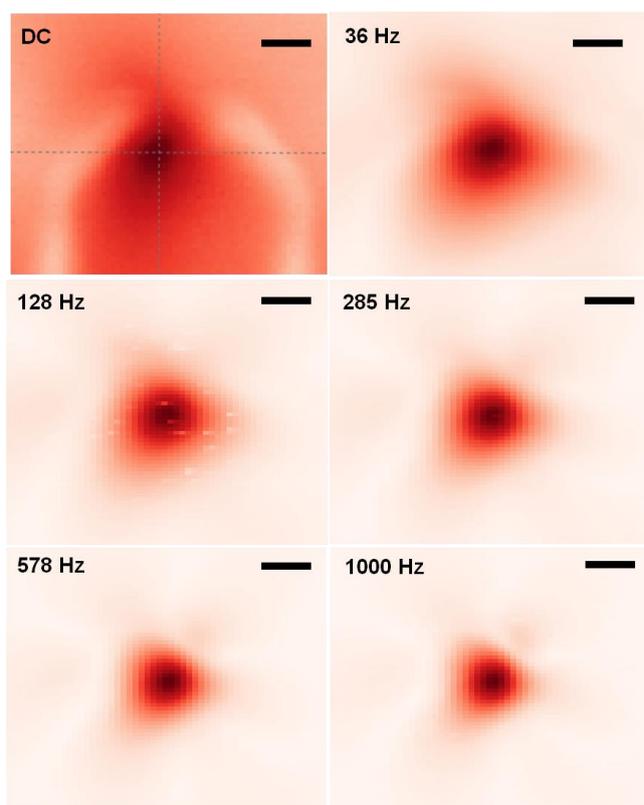

**Figure S5.1** Normalized responsivity maps of the same photodetector at different frequencies. Scale bars are 100 µm.

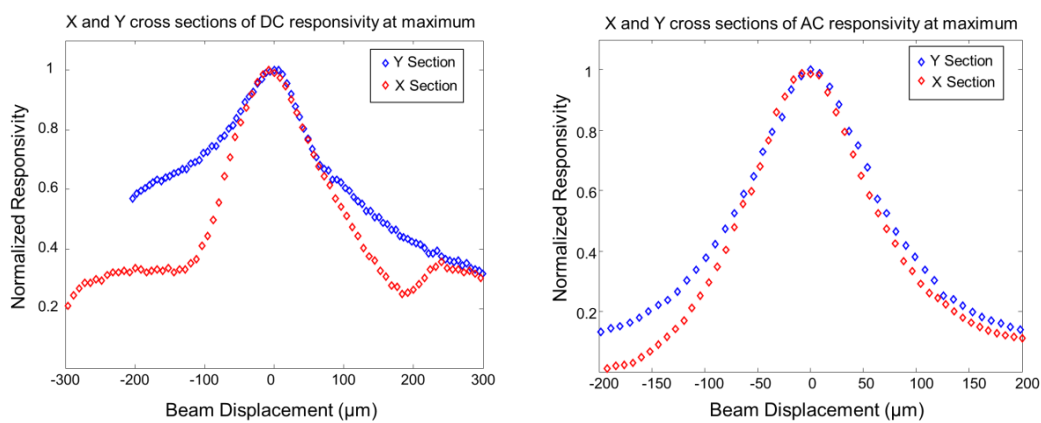

**Figure S5.2** Cross sections of normalized DC and AC (77 Hz) responsivity along principal axes passing through the maximum (along axes marked in fig. S2)

*S.6 Model for photodetector response of slabs of pyroelectric materials with graphene*

In order to study the heat transfer and thermal characteristics of the pyro-resistive photodector presented in this work, we can use a one dimensional model that will give insights for understanding the device.

*S6.1 One-dimensional model*

The schematics of the device is sketched in fig.S6, where we assume heat propagation for the system to be one-dimensional along the vertical direction, z. The system is formed by a slab of LiNbO$_3$ and a sheet of graphene deposited on top. In such case, the heat propagation is modeled by the following equation:

$$\frac{\partial^2 T}{\partial z^2} - \xi^2 T = \frac{1}{k} g(z, \omega) \qquad \text{SI.13}$$

Where $T$ is the temperature increase with respect to the environment, $\xi^2 = j\omega/\alpha$ is the so-called fin term including the angular frequency ω and the substrate's thermal diffusivity α, g(z,ω) is the periodic heat introduced by the laser absorption. Together with equation SI.13 we need to specify the boundary conditions at z=0 and z=d. In our case the boundary condition will have the general form:

$$k \frac{\partial T}{\partial z} + [h_i + j\omega(\rho cb)_i] T = f_i(z, \omega) \qquad \text{SI.14}$$

Those general boundary conditions accounts for convective heat transfer via h and thermal dissipation through a thin film of thickness b with density ρ and heat capacitance c. If we indicate the Green's function solution G the steady-periodic temperature is given by:

$$T(z, \omega) = \frac{\alpha}{k} \int g(\bar{z}, \omega) G(z, \bar{z}, \omega) d\bar{z} + \frac{\alpha}{k} \sum_i \int f_i(\bar{z}, \omega) G(z, \bar{z}, \omega) d\bar{z} \qquad \text{SI.15}$$

Where the general solution Green's function is:

$$G(z,\bar{z};\xi) = \frac{S_2^-(S_1^- e^{-\xi(2L-|z-\bar{z}|)}+S_1^+ e^{-\xi(2L-z-\bar{z})})}{2\alpha\xi(S_1^+ S_2^+ - S_1^- S_2^- e^{-2\xi L})} + \frac{S_2^+(S_1^+ e^{-\xi(|z-\bar{z}|)}+S_1^- e^{-\xi(z+\bar{z})})}{2\alpha\xi(S_1^+ S_2^+ - S_1^- S_2^- e^{-2\xi L})} \qquad \text{SI.16}$$

$$S_i^\pm = k\xi \pm \lambda_i \qquad \text{SI.17}$$

where $\xi^2 = j\omega/\alpha$, and i = 1 corresponds to z = 0 and i = 2 to z = L, $\lambda_i = h_i + j\omega(\rho cb)_i = h_i + \alpha\xi^2 k_i^F$ accounts for convection of air ($h_i$) and dissipation of heat in a thin film of thickness b.

*S6.2 Thick slab of LiNbO3 with absorption of light through the substrate*

The device presented in the paper, consisted of a 500µm -thick slab of $LiNbO_3$ with graphene on top of it. At the opposite face where graphene was deposited, the face was covered with a layer of Cr/Au of thickness 5/100nm in order to electrically ground the face.

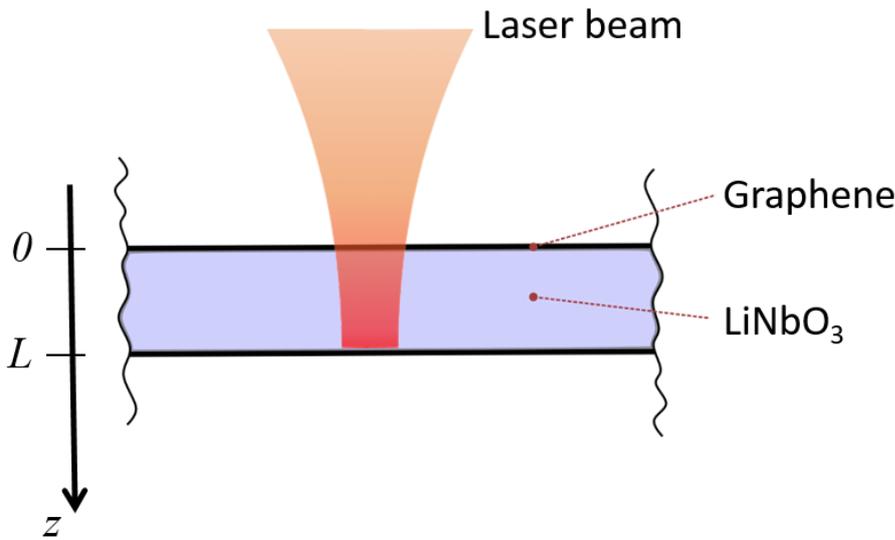

**Figure S6.1** Schematics of pyroresistive detector based on a thin membrane of $LiNbO_3$ comprising: the $LiNbO_3$ substrate, graphene on top of it, and a buffer layer to isolate graphene from the topmost IR absorption layer.

Thus, the boundary conditions SI.13 become:

$$k\frac{\partial T}{\partial z}\bigg|_{z=0} + h_{Air}T = 0 \qquad \text{SI.18}$$

$$k\frac{\partial T}{\partial z}\bigg|_{z=L} + j\omega(\rho cb)_{Au}T = 0 \qquad \text{SI.19}$$

If we restate the last boundary condition in terms of $j\omega = \alpha\xi$ and $k_i^F = (\rho cb)_{Au}$, this leads to the following relations for the S parameters in the Green's function SI. 16

$$S_1^- = k\xi + h_{Air}$$
$$S_1^+ = k\xi - h_{Air} \qquad \text{SI.20}$$
$$S_2^- = k\xi + \alpha\xi^2 k_i^F$$
$$S_2^+ = k\xi - \alpha\xi^2 k_i^F$$

The temperature profile $T(z;\xi)$ at $z=0$ can be obtained in the slab upon heating with a source $g(z)$ having this form:

$$g(z) = g_0[e^{-az} + Re^{-a(2L-z)}] \qquad \text{SI.21}$$

where $g_0$ is the heat maximum generated upon laser illumiation and R is the optical reflection coefficient on the face $z=L$. Thus the temperature $T(z=0;\xi)$ is given by:

$$T(0;\xi) = \frac{a}{k}\int_0^L G(z,\bar{z};\xi)g(\bar{z})d\bar{z} = \frac{1}{\xi^2-a^2}\frac{g_0 e^{-2aL}(S_1^-+S_1^+)}{2k\xi(S_1^-S_2^- - e^{2L\xi}S_1^+S_2^+)}[ae^{2aL}S_2^- - aRS_2^- - ae^{2L\xi}RS_2^+ + e^{2L(a+\xi)}S_2^+(a-\xi) + e^{2aL}S_2^-\xi + RS_2^-\xi - e^{2L\mu}RS_2^+\xi + e^{L(a+\xi)}(a(-1+R)(S_2^- + S_2^+) - (1+R)(S_2^- + S_2^+)\xi] \qquad \text{SI.22}$$

The response in DC ($\omega = 0$ implies $\xi = 0$ is $T_{DC} = T_{DC} = T(0;\xi=0)$. If we take the amplitude of the normalized response $H(\omega) = T(0;\xi)/T_{DC}$ and express it in terms of $\omega$ we can write:

$$H(\omega) = \frac{|T(0;\omega)|}{|T(0;\omega=0)|} = \left|\frac{1}{a^2+j\omega/\alpha}\right||F(\omega)| = \frac{1}{\sqrt{1+\omega^2\tau^2}}|\widehat{F}(\omega)| \qquad \text{SI.23}$$

where $\tau = 1/\alpha a^2$ and $\widehat{F}(\omega)$ can be approximated by a polynomial rational function $\sum a_n \omega^n / \sum b_n \omega^n$ which is convenient in order to fit experimental data rather than a complex expression like the one in eq.SI.22 with so many physical parameters. In the manuscript n=2 was chosen in the rational polynomials and gave good results.

*S6.3 Thin slab of pyroelectric material with absorption layer*

In order to evaluate the potential of the pyroresistive detector, we consider a slab of pyroelectric material as $LiNbO_3$ with a thickness L, covered with a graphene sheet that will act as the sensing element, an insulating buffer layer (e.g. $Al_2O_3$) on top and a thin film as an absorption layer (e.g. Nichrome, Si) on the topmost layer. The buffer layer needs to be sufficiently thick to guarantee electrical insulation with respect to the absorption layer, so a few tenths of nm would be enough and from a thermal propagation point of view it won't have a significant impact. If needed it could be accounted for in the model.

The absorption layer absorbs at least 90% of the incoming light and can be also used as an integrated spectral filter by patterning or selecting the appropriate materials. The absorbed light that is periodically modulated will be converted into heat that propagates into $LiNbO_3$ thus generating the pyroresistive effect on graphene.

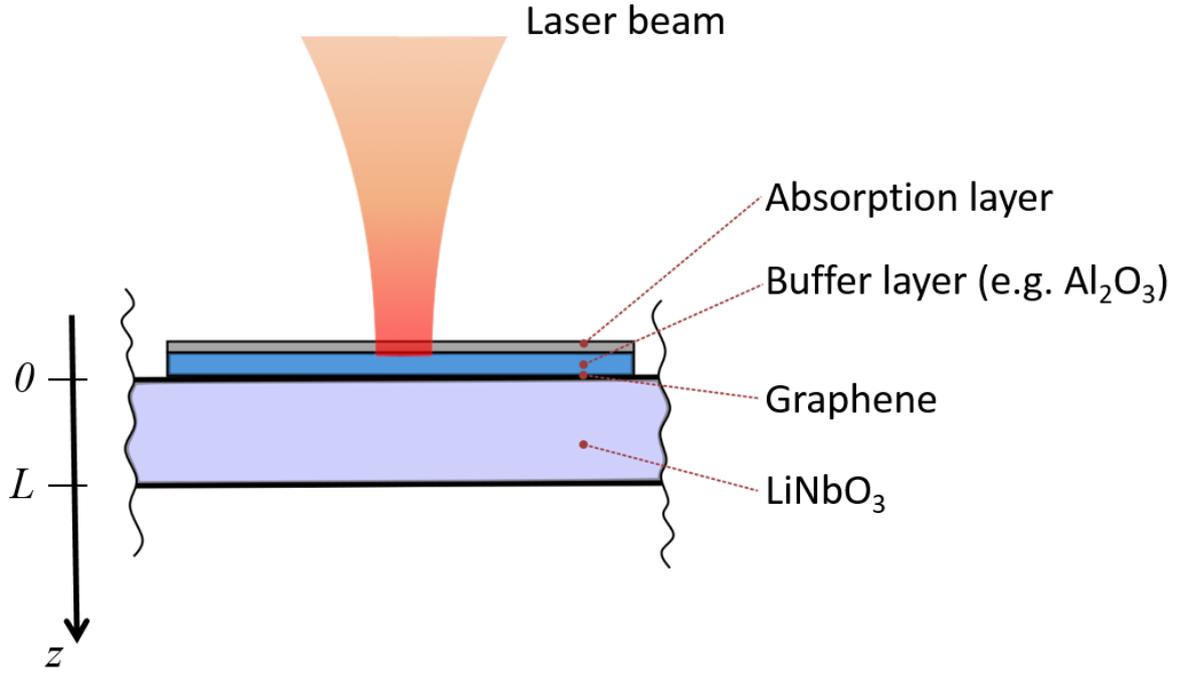

**Figure S6.2:** Schematics of pyroresistive detector based on a thin membrane of LiNbO$_3$ comprising: the LiNbO$_3$ substrate, graphene on top of it, and a buffer layer to isolate graphene from the topmost IR absorption layer.

Equation SI.13 holds for the unidimensional structure considered, with the following boundary conditions at the top and bottom sides:

$$-k \frac{\partial T}{\partial z}\bigg|_{z=0} = h_1 T + q_0(\omega) \quad \text{SI.24}$$

$$-k \frac{\partial T}{\partial z}\bigg|_{z=L} = h_2 T \quad \text{SI.25}$$

where $k$ is the thermal conduction coefficient for LiNbO$_3$, h is the convection coefficient of the surrounding medium (e.g. air) and q$_0$ is the intensity of heat generated at the top layer by absorption of light. The solution at z=0 is:

$$|T(0; \mu| = \left| \frac{g_0(k\xi \cosh(\xi L) + h_2 \sinh(\xi L))}{(h_1+h_2)k\xi \cosh(\xi L) + (h_1 h_2 + k^2\xi^2)\sinh(\xi L)} \right| \quad \text{SI.26}$$

Where $\xi = \sqrt{i\omega/\alpha}$

In figure S6 we plot the comparison between the temperature increase per unit of optical power for the different models presented in S6.2 and S6.3. The placement of a thin layer of 300nm of Nichrome and 40nm of $Al_2O_3$ on top of graphene allows to improve the performance of three orders of magnitude according to the model. Although the model is relatively simple and only one-dimensional, it gives much room for improvement in the expected responsivity for the pyro-resistive photodetector. Moreover, we have an increase in the bandwidth to tenths of Hz, allowing the pyroresistive photodetectors to potentially reach a response time of the order of ms.

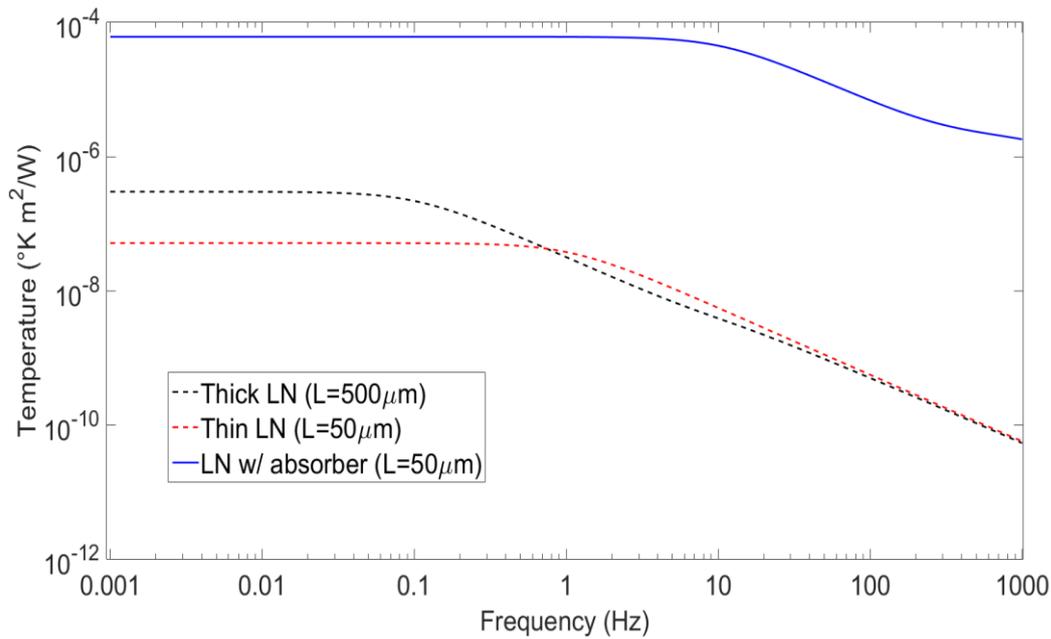

**Figure S6.3** Comparison of models of pyroresistive photodetector for different thickness and with an Nichrome IR absorber layer on top of graphene.

## S.7 Details of experimental methods

*Device fabrication:*

The devices were fabricated on the +Z face of 500 μm thick LiNbO$_3$ substrates (MTI Corporation). The substrates were cleaned using organic solvents followed by diluted basic piranha solution to remove any impurities that might prevent good adhesion between graphene and the substrate. To avoid thermal shock and discharge of static charges (due to the pyroelectric nature of the substrate), all baking processes were done with a slow ramping of temperature. UV lithography was used to define two point and 4 point (Hall geometry) devices on the substrate. Subsequently, Cr (5nm)/Au (100nm) contacts were deposited by thermal evaporation and lifted off using acetone. Additionally, the rear side of all substrates were coated with Cr (5nm)/Au (100nm) to aid discharge of static charges and act as a thermal sink. CVD grown monolayer graphene (Graphenea Inc.) on Cu foil was coated with PMMA (100 nm). Wet etching of the Cu foil was done using Ammonium persulfate solution for about 5 hours. The PMMA on graphene film was then rinsed with DI water and lifted up with the LiNbO3 substrate. The sample was then left overnight in vacuum to promote good adhesion. After PMMA removal, the graphene pattern was defined by UV lithography and using Ar/O$_2$ plasma at 10W for about 1 minute.

*Top gating measurements:*

For top gating measurements, Polyethylene oxide was dispersed with LiClO$_4$.3H$_2$O (with a ratio of 2:1 approximately) in methanol, and then drop-cast on the graphene device. Once methanol is evaporated the polymer acts as a solid solvent for ions (ionic gel). Ions accumulate in the vicinity of graphene when a potential voltage is applied between the device and an external electrode. This allows to use it as very efficient local top gate transparent to most wavelengths.[2]

*Photoresponse measurements:*

For the AC photoresponse mapping, the sample is mounted on a motorized stage and moved across a focussed beam from a laser. The FWHM of the spot size is comparable to the wavelength. The laser light is modulated between 1-1000 Hz using an optical chopper. The photocurrent is amplified by a Femto DLPCA-200 preamplifier and the lock-in signal is obtained by a Stanford Research Systems SR830 DSP. DC photocurrent map have been taken in a similar way as the AC ones for the device scanning, with continuous (unchopped) light and using a 1Hz low-pass filter after the current amplification. All measurements are taken in a $N_2$ atmosphere.

*S.8 Thermal length analysis*

To extract the thermal length associated with heat propagation and its frequency dependence we analyzed the cross sections of the normalized photoresponse maps taken at different chopper frequencies (Fig. S5.1). The cross sections were taken at the maximum values of the photoresponse along the y axis. In figure S8.1 we report the data extracted where only one side is considered for the sake of clarity and the value of the maximum responsivity corresponds to zero. The cross sections present three different regimes depending on the distance from the zero value, which can be fitted with proper functions as shown on figure S8.1.

The region near zero presents a behavior very close to Gaussian (red solid curves in figure S5.1). For this region, which is of high interest in applications and device designs, we can define the half-width-at-half maximum of the gaussian photoresponse distribution as the thermal length. The thermal length and the fitting by a power law as $Af^{\Box}$ are presented and discussed in the main text (see figure 3b).

Moving further from the peak of the photoresponse the behavior seems to follow an exponential decay with two different decay rates (green and blue solid curves in figure S5.1). The functions used to fit these the two regions with different decays are $\sim\exp(-\Box x)$ where the coefficients $\Box\Box\Box\Box f\Box\Box$ depends on the frequency $f$. The values of the coefficients $\Box$ calculated from the fittings are reported in figure S5.2.

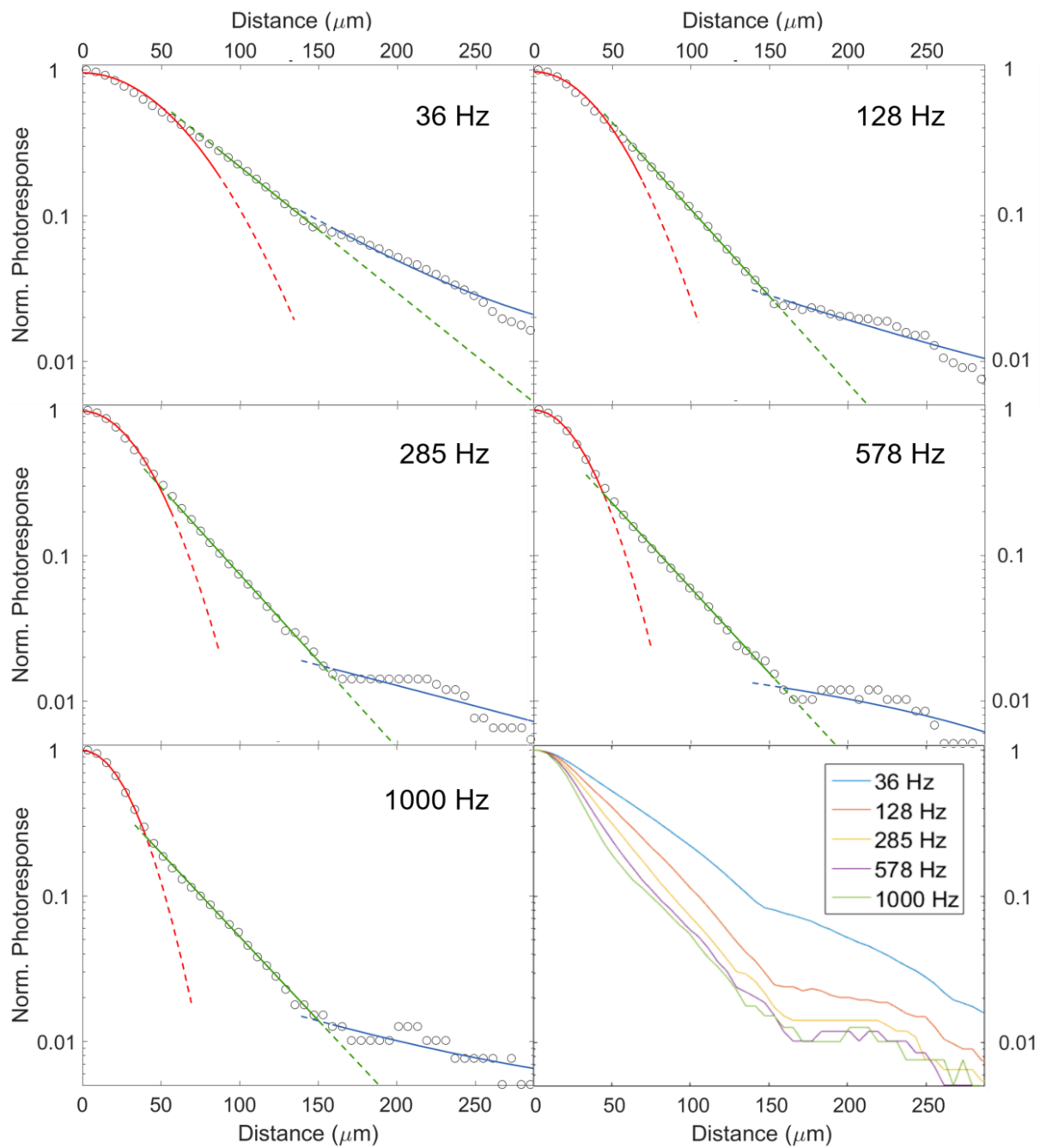

**Figure S8.1** Data and fitting for the cross sections of the photoresponse maps at different chopping frequencies. The data suggest three different radial decay regimes starting from the point of maximum photoresponse: (i) Gaussian (red curves) close to the maximum, from which we can extract the thermal length, followed by (ii) and (iii) exponential decays with two different rates (green and blue curves).

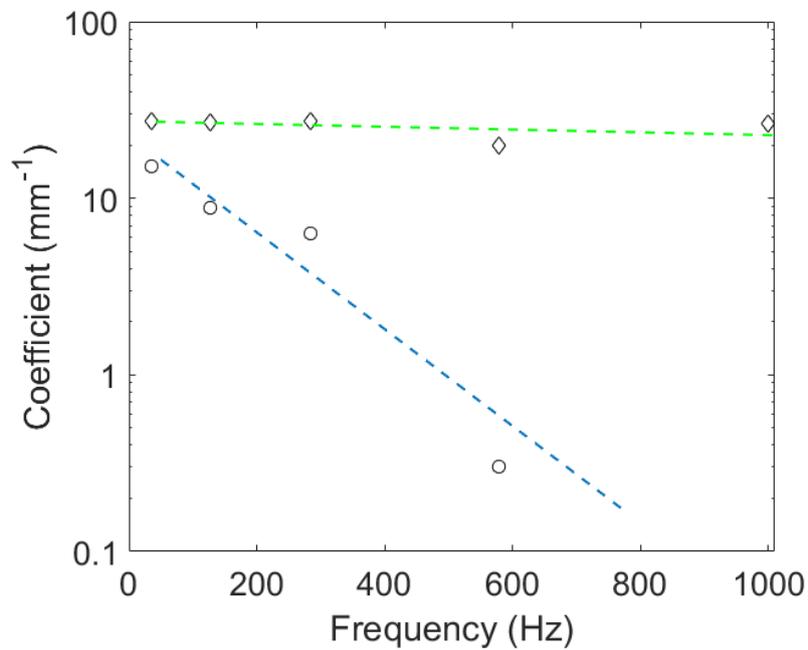

**Figure S8.2** Comparison of the two longer decays of photoresponse. Diamonds are coefficient from the fitting of green curves and circles are from blue curves in figure S8.1. The dotted lines serve as a guide to the eye.

References:

[1] Y. V. Shaldin, V. T. Gabriélyan, S. Matyjasik, Crystallography Reports 2008, 53, 847.

[2] C.-F. Chen, C.-H. Park, B. W. Boudouris, J. Horng, B. Geng, C. Girit, A. Zettl, M. F. Crommie, R. A. Segalman, S. G. Louie, F. Wang, Nature 2011, 471, 617.